\documentclass[annual]{acmsiggraph}

\usepackage[margin=1in]{geometry}

\title{Green's functions for non-classical transport with general anisotropic scattering}

\author{ Eugene d'Eon \\
NVIDIA }

\pdfauthor{ Eugene d'Eon  }


\usepackage{smrdefaults}

\usepackage{color}

\usepackage{multirow}

\usepackage{rotating}
\newcommand{\s}{\langle s_c \rangle}

\newcommand{\e}{\text{e}}

\newcommand{\pos}{\mathbf{x}}

\newcommand{\dir}{\Omega}

\def\rm{R^{-}}

\usepackage{paralist}
\usepackage{subfigure}
\usepackage{rotating}
\usepackage{amsgen}
 
\usepackage{alltt}
\usepackage{ulem}
\usepackage{xcolor}

\begin{document}

\normalem

\maketitle

\begin{abstract}
  In non-classical linear transport the chord length distribution between collisions is non-exponential and attenuation does not respect Beer's law.  Generalized radiative transfer (GRT) extends the classical theory to account for such two-point correlation between collisions and neglects all higher order correlations.  For this form of transport, we derive the exact time-independent Green's functions for the isotropic point source in infinite 3D homogeneous media with general anisotropic scattering.  Green's functions for both collision rate density, which characterizes absorption and reaction rates in the system, and radiance/flux, which characterizes displacement of radiation/particles, are solved in Fourier space.  We validate the derivations using gamma random flights to produce the first anisotropic scattering benchmark solutions for the generalized linear Boltzmann equation.  For gamma-4 flights with linearly anisotropic scattering and gamma-6 flights with Rayleigh scattering the collision rate density is found explicitly in real space as a sum of diffusion modes.
\end{abstract}

\keywordlist

\section{Introduction}

  Linear transport of monoenergetic neutral particles in stochatic geometries can significantly deviate from the predictions of classical transport theory~\cite{chandrasekhar60,davison57} when there are significant spatial correlations in the geometries.  This has motivated the use of non-exponential random flights to form generalized radiative transfer (GRT) theories that exactly exhibit some desired non-exponential distribution of chord lengths between collisions~\cite{burrus1960radiation,doub1961particle,randall1964,rybicki1965transfer,alt1980biased,sahni1989equivalence,audic1993monte,kostinski01,davis04,davis06,moon07,taine2010generalized,frank10,larsen11,zarrouati2013statistical,vasques13,davis14,frank15,xu16,rukolaine2016generalized,liemert2017radiative,binzoni2018generalized,frank2018fractional,deon2018reciprocal,jarabo18,bitterli2018radiative}.  We derive new green's functions for these theories in the case of anisotropic scattering.
  
  Point source Green's functions for monoenergetic transport in infinite homogeneous medium with a completely general symmetric phase function are known exactly in classical exponential media via Fourier transform~\cite{davison00,wallace1948angular,grosjean51,vanmassenhove67,williams1977role,paasschens97,ganapol03,zoia11e} and have been studied using other approaches, such as path integrals~\cite{tessendorf11}.   These functions have a number of direct applications~\cite{narasimhan2003shedding}, they provide important benchmarks for checking correctness of more advanced codes~\cite{ganapol08} and they play a role in bounded media via Placzek's lemma~\cite{case53}.  The goal of this paper is to derive such Green's functions for GRT.

  In the case of isotropic scattering, the Fourier transform approach has been extended for GRT and the Green's functions are known~\cite{deon14,deon2019reciprocalii}.  In this paper, we show how to use the derivations of Grosjean~\shortcite{grosjean51} to derive the exact forms for general anisotropic scattering.  To the best of our knowledge, Grosjean's 1951 derivation has never been validated numerically in the non-exponential case with anisotropic scattering.  We perform Monte Carlo simulation of gamma random flights in 3D and find good agreement.  We omit the lengthy details of Grosjean's full derivation and refer the interested reader to his monograph.  Our present goal is to briefly highlight only those details that are required to form benchmark solutions in GRT and discuss their role in these theories.

  \section{General Theory}

    \subsection{The random flights of GRT}
\begin{figure}
        \centering
        \subfigure[]{\includegraphics[width=.49\linewidth]{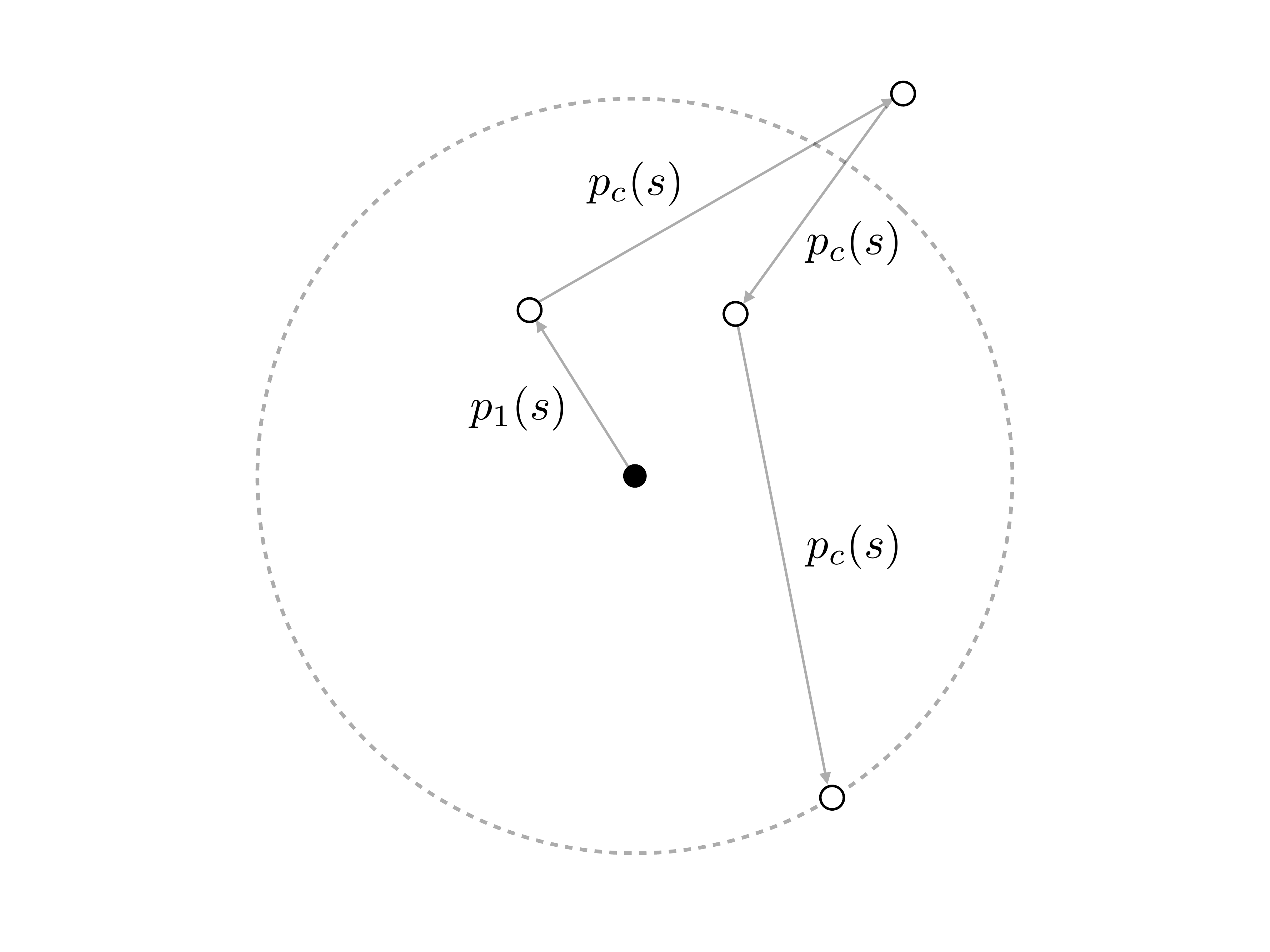}}
        \subfigure[]{\includegraphics[width=.49\linewidth]{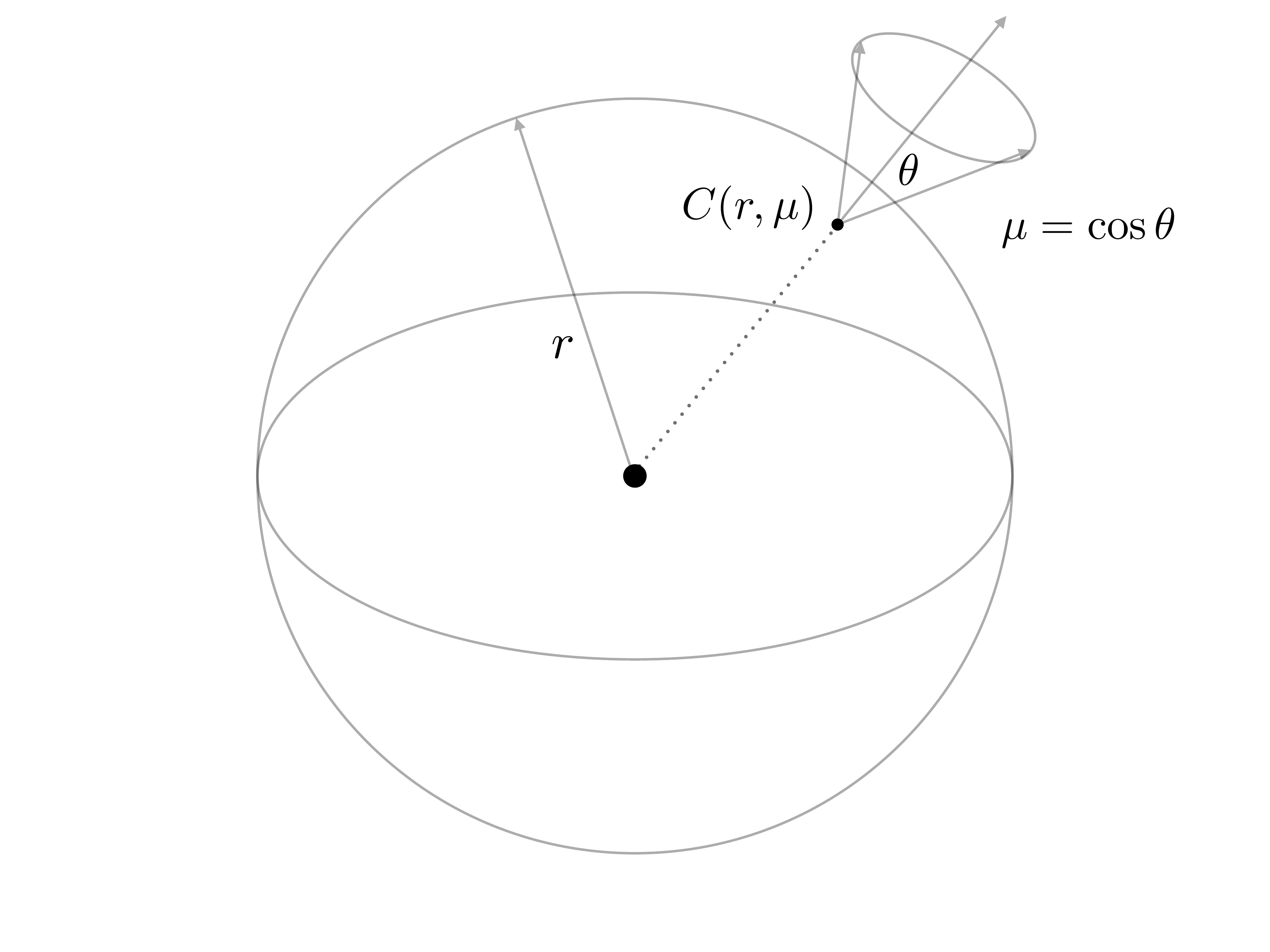}}
        \caption{(a) We consider the random flight where a particle leaves an isotropic point source in 3D (black circle) and travels at constant speed along straight paths between collision events (white cirlces).  The free-path lengths between collisions are random variates drawn from the distribution $p_c(s)$.  The initial free-path length $p_1(s)$ may be different from $p_c(s)$.  At each collision the particle scatters according to a given phase function $P(\Omega_i \rightarrow \Omega_o)$ with probability $c$ and is absorbed otherwise, terminating the flight.  (b) We seek the rate density $C(r,\mu)$ that particles enter collisions at some distance $r$ from the point source restricted to a cone of directions indexed by cosine $\mu$ to the position vector.}
        \label{fig:sphere_geom} 
      \end{figure}

      Let us consider a unit isotropic point source at the origin of a homogeneous infinite 3D medium emitting particles that travel at constant speed $v$ along piecewise straight paths (Figure~\ref{fig:sphere_geom}).  Collisions between the particle and the medium result either in absorption, with probability $1 - c$, or scattering into a new direction using a given phase function with probability $0 < c \leq 1$.  We consider a general symmetric phase function with Legendre expansion
      \begin{equation}\label{eq:phaseexpansion}
        P(\Omega_i \rightarrow \Omega_o)  = \frac{1}{4\pi} \sum_{l=0}^\infty (2l+1) f_l P_l(\Omega_i \cdot \Omega_o) = \frac{1}{4\pi} \sum_{l=0}^\infty A_l P_l(\Omega_i \cdot \Omega_o)
      \end{equation}
      where $\Omega_i$ is the direction before collision and $\Omega_o$ is the scattered direction.  The phase function is normalized over the unit sphere, $A_0 = 1$.

      Free path-lengths between collisions are drawn from a given normalized distribution $p_c(s)$, which can be estimated by Monte Carlo tracking in specific realizations of the stochastic geometry~\cite{audic1993monte,moon07,larsen11} or derived from a known attenuation law in the system~\cite{torquato93,deon2018reciprocal}.  We denote the initial free-path length distribution from the point source $p_1(s)$, which is either $p_c(s)$ or a related distribution, $p_u(s)$.  If emission is spatially correlated to the first scattering event in the same way that spatial correlation arises between collisions then $p_1(s) = p_c(s)$ and the full random flight is homogeneous.  However, emission from a deterministic location in the stochastic geometry must consider statistics to the next collision using all realizations of the geometry and not only those that have a particle at the origin from which to begin the flight.  For this reason, emission from a so called uncorrelated origin has free-path length statistics $p_u(s)$ that are distinct from the intercollision statistics $p_c(s)$~\cite{audic1993monte,deon2018reciprocal}.  An equivalent distinction has been recognized for continuous-time random walks~\cite{feller1971introduction,tunaley1974theory,tunaley1976moments,weissrubin1983random} where it was proposed that\footnote{Specifically, this occurs in the case of tightly-coupled walks where the wait time is proportional to the displacement, which corresponds to the constant speed model.}
      \begin{equation}\label{eq:pcpu}
        p_c(s) = -p_u'(s) / \s, \quad \s \equiv \int_0^\infty p_c(s) s ds
      \end{equation}
      where $\s$ is the mean free path length between collisions.  Eq.(\ref{eq:pcpu}) was also found to be a necessary condition for achieving Helmholtz reciprocity in GRT~\cite{deon2018reciprocal}, which is desired given that reciprocity is observed in any realization of the stochastic geometry and therefore expected of the mean transport.

      \subsubsection{Collision-rate density}
      Now consider a single particle leaving the isotropic point source.  Our primary quantity of interest in GRT is the collision rate density, since most reaction rates in a given system will be proportional to this quantity.  The \emph{angular collision-rate density} $C(r,\mu)$ is defined such that $4 \pi r^2 C(r,\mu) dr d\mu$ is the mean number of collisions (either scattering or absorbing) experienced by the particle within a shell of radii $[r,r+dr]$ about the point source who directions \emph{before} collision have cosines in $[\mu, \mu+d\mu]$.  Direction cosines $\mu$ are measured with respect to the normalized position vector and index cones of directions with $\mu = 1$ pointing away from the point source (Figure~\ref{fig:sphere_geom}).  We denote the scalar collision rate density by
      \begin{equation}
        C(r) = \int_{-1}^1 C(r,\mu) d\mu.
      \end{equation}
      To distinguish between correlated and uncorrelated emission we also write $C_c(r)$ and $C_u(r)$, respectively for the scalar collision rate densities.
      With non-stochastic single-scattering albedo $c$ at every collision, the mean number of collisions is $1 / (1-c)$ and therefore the 0th moment of $C$ is~\cite{ivanov1994resolvent}
      \begin{equation}
        \int_0^\infty 4 \pi r^2 C(r) dr = \frac{1}{1-c},
      \end{equation}
      for all phase functions and free-path length distributions.  In a system with spherical symmetry, which will always be the case in this paper, the angular collision rate density $C(\pos,\dir)$ at position $\pos$ and direction $\dir$ is related to $C(r,\mu)$ by $C(\pos,\dir) = C(r,\mu) / (2\pi)$.

      The phase function and single-scattering albedo immediately give the distribution of particles in the system as they leave collisions.  The in-scattering rate density $B(\pos,\dir)$ is related to the collision rate density by~\cite{larsen11}
      \begin{equation}
        B(\pos,\dir) = c \int_{4\pi} P(\dir' \cdot \dir) C(\pos,\dir') d\dir'
      \end{equation}
      and satisfies the generalized Peierls integral equation,
      \begin{equation}\label{eq:integralB}
        B(\pos,\dir) = c \iiint P\left(\Omega \cdot \frac{\pos - \pos'}{|\pos - \pos'|}\right) \left\{ \left[ B\left(\pos',\frac{\pos - \pos'}{|\pos - \pos'|}\right) + \frac{Q_c(\pos')}{4 \pi} \right] \frac{p_c(|\pos-\pos'|)}{|\pos-\pos'|^2} + \frac{Q_u(\pos')}{4 \pi} \frac{p_u(|\pos-\pos'|)}{|\pos-\pos'|^2} \right\} dV'.
      \end{equation}
      This is an integral equation of non-homogeneous non-exponential random flights~\cite{grosjean51,weissrubin1983random} and is the basis of our derivation for Green's functions in GRT.  
      
      \subsubsection{Radiance and fluence}
      We are also interested in finding Green's functions for the radiance and fluence in the system.  In GRT, the semi-Markov nature of the transport breaks the classical proportionality between collision rate and radiance~\cite{deon2019reciprocalii}, and the two are no longer related by the inverse mean free path, requiring separate Green's function derivations and separate Monte Carlo estimators.  The radiance in GRT satisfies a more complicated equation, the generalized linear Boltzmann equation (GLBE).  However, the GLBE has been proven equivalent to Eq.(\ref{eq:I})~\cite{larsen11}, and the radiance $I(\pos,\dir)$ (specific intensity/angular flux) is simply related to $B$ by~\cite{larsen11}
      \begin{equation}\label{eq:I}
        I(\pos,\dir) = \int_0^\infty \left[ B(\pos-s \dir,\dir) + \frac{Q_c(\pos-s \dir)}{4 \pi} \right] X_c(s) + \frac{Q_u(\pos-s \dir)}{4 \pi} X_u(s) ds
      \end{equation}
      where
      \begin{equation}\label{eq:Xc}
        X_c(s) = \int_s^\infty p_c(s') ds'
      \end{equation}
      is the attenuation law leaving a collision~\cite{larsen11} and
      \begin{equation}
        X_u(s) = \int_s^\infty p_u(s') ds'
      \end{equation}
      is the uncorrelated-origin attenuation law~\cite{deon2018reciprocal}.  Finally, the fluence (scalar flux) is
      \begin{equation}\label{eq:fluence}
        \phi(\pos) = \int_{4\pi} I(\pos,\dir) d \dir.
      \end{equation}
      For a single point source, where either one of $Q_c$ or $Q_u$ is a unit dirac delta at the origin and the other is zero, it suffices to solve Eq.(\ref{eq:integralB}) for $B(r,\mu)$, and then the radiance and fluence are determined after by Eq.(\ref{eq:I}).  We solve for these Green's functions in the next section using Fourier transforms.
      
      \subsubsection{Relationship to previous GRT models}
      This formulation of random flights is an extension of the generalized models of Larsen~\shortcite{larsen11} and Davis~\shortcite{davis06} to support reciprocal transport in bounded domains by including $p_u(s)$, following \cite{audic1993monte}, who proposed such an extension for Markovian binary mixtures.  Boundary conditions for this extension were given in an earlier paper~\cite{deon2019reciprocalii}.  The extended integral equations Eq.(\ref{eq:integralB}) and (\ref{eq:I}) are also analagous to Sahni's~\shortcite{sahni1989equivalence}, where he considered a different class of Markovian binary mixtures than Audic and Frisch in what could be considered the first proposal of a non-exponential two-point transport theory.  Our extended formulation also supports several proposed forms of non-exponential random flights where all free paths use $p_u(s)$~\cite{davis06,taine2010generalized,davis14,wrenninge17,liemert2017radiative,binzoni2018generalized}.  However, application of these models and related Green's functions to bounded domains are known to result in non-reciprocal transport.
 
    \subsection{Grosjean's solution}
  
    In his thesis, Grosjean~\shortcite{grosjean51} solved a fully general random flight in an infinite 3D medium with an isotropic point source.  He solved for $C(r,\mu)$ and $B(r,\mu)$, given in spherical harmonic expansions whose coefficient functions are given by Fourier inversion (requiring numerical inversion in most cases).  In the most general form of his work, Grosjean permitted free-path-length distributions between collisions $p_n(s)$, single-scattering albedos $c_n(s)$, and scattering kernels (phase functions) $P_n(\Omega_i \rightarrow \Omega_o)$ that are chosen independently for each collision order $n > 0$, to build fully heterogeneous random flights.  He also presented simplifications for the case of a completely homogeneous random flight where the intercollision free-path length distribution $p_c(s)$ and phase function $P(\Omega_i \rightarrow \Omega_o)$ are identical for all collision orders $n$ and the single-scattering albedo $0 < c \leq 1$ at every collision is the same constant.

    For our homogeneous random flight ($p_1(s) = p_c(s)$), Grosjean showed that the angular collision rate density has a spherical harmonic expansion given by
      \begin{equation}\label{eq:C}
        C(r,\mu) = \sum_{l=0}^\infty C_l(r) (2l+1) P_l(\mu)
      \end{equation}
    where
      \begin{equation}
        C_l(r) = \frac{1}{8 \pi c} \int_0^\infty h^{(l)}(u) \, u \, j_l(r \, u) du
      \end{equation}
    in terms of expansion functions $h^{(l)}(u)$ that are the solution of the linear system (\cite{grosjean51}, p. 77)
      \begin{equation}\label{eq:hsystem}
        h^{(l)}(u) = \frac{2}{\pi} u F^{(l,0)}(u) + \sum_{m=0}^\infty A_m h^{(m)}(u) F^{(l,m)}(u).
      \end{equation}
    The functions $F$ are given by (\cite{grosjean51}, p. 70)
      \begin{align}\label{eq:F}
        F^{(l,m)}(u) &= \frac{c}{2} i^{m-l} \int_0^\infty \int_{-1}^1 p_c(z)  P_l(\mu ) P_m(\mu ) e^{i \mu  u z} \, d\mu \, dz \\&= c \int_0^\infty  p_c(y) dy \left[ i^m P_m\left( \frac{d}{i d z} \right)\left( j_l(z) \right) \right]_{z \equiv y u}
      \end{align}
    where $j_l(z)$ is the spherical Bessel function
      \begin{equation}
        j_l(z) = \frac{\sqrt{\frac{\pi }{2}} J_{l+1/2}(z)}{\sqrt{z}}.
      \end{equation}
    The notation
      \begin{equation}
        P_m\left( \frac{d}{i d z} \right)\left( j_l(z) \right)
      \end{equation}
    is understood to mean the differential operator formed from replacing $z^n$ in the expansion of Legendre polynomial $P_m(z)$ with
      \begin{equation}
        \frac{\partial^n}{i^n \partial z^n}
      \end{equation}
    applied to $j_l(z)$.  The functions $F$ obey a symmetry $F^{(l,m)}(u) = (-1)^{(l+m)} F^{(m,l)}(u)$.

    The Fourier integrals for $C_l(r)$ may only be convergent in the sense of Cesaro summability, prompting the separation of the density of first collisions to express the total collision-rate density as (\cite{grosjean51}, p.75)
      \begin{equation}\label{eq:Crmuplus}
        C(r,\mu) = \frac{p_c(r) }{4 \pi r^2} \delta(1-\mu) + \sum_{l=0}^\infty C_l^+(r) (2l+1) P_l(\mu)
      \end{equation}
    with
      \begin{equation}
        C_l^+(r) = \frac{1}{8 \pi c} \int_0^\infty \left( h^{(l)}(u) - \frac{2 u}{\pi} F^{(l,0)} \right) \, u \, j_l(r \, u) du.
      \end{equation}
    
    From these derivations we immediately have the scalar collision-rate density $C(r) = 2 C_0(r)$ for correlated emission.  
    
    For the case of uncorrelated emission, $p_1(s) = p_u(s)$, and we refer to Grosjean's more general derivation.  For this almost homogeneous random flight, we find a modified system of equations for the expansion functions $h$ in Eq.(\ref{eq:C})
      \begin{equation}\label{eq:hsystemU}
        h^{(l)}(u) = \frac{2}{\pi} u F_1^{(l,0)}(u) + \sum_{m=0}^\infty A_m h^{(m)}(u) F^{(l,m)}(u)
      \end{equation}
    where the functions $F_1$ arise from appropriately modifying Eq.(\ref{eq:F}) to include the free-path length distribution $p_1(s)$ instead of always $p_c(s)$, giving
      \begin{equation}\label{eq:F1}
        F_1^{(l,m)}(u) = c \int_0^\infty  p_1(y) dy \left[ i^m P_m\left( \frac{d}{i d z} \right)\left( j_l(z) \right) \right]_{z \equiv y u}.
      \end{equation}
  
    For the radiance and fluence, we use Grosjean's solutions for the Neumann series of a heterogeneous flight (\cite{grosjean51}, Eqs.(259,259')).  The rate density for the particle to enter its nth collision at r with cosine $\mu$ is
    \begin{equation}\label{eq:C}
      C(r,\mu|n) = \sum_{l=0}^\infty C_l(r|n) (2l+1) P_l(\mu)
    \end{equation}
    where
    \begin{equation}
      C_l(r|n) = \frac{1}{8 \pi c} \int_0^\infty h_n^{(l)}(u) \, u \, j_l(r \, u) du
    \end{equation}
    and the Neumann series $h$ functions are 
    \begin{align}\label{eq:hseries}
      &h_n^{(l)}(u) = \sum_{m=0}^\infty A_m h_{n-1} F_n^{(l,m)}, \quad (n = 2, 3, 4...) \\
      &h_1^{(l)}(u) = 2 u F_1^{(l,0)}
    \end{align}
    where $ F_n^{(l,m)}$ are defined using the free-path length distribution for the nth free path $p_n(s)$.  The homogeneous system of equations in Eq.(\ref{eq:hsystem}) follows from Eqs.(\ref{eq:hseries}) using the definition $h^{(l)}(u) = \sum_{n=1}^\infty h_n^{(l)}(u)$ and that $F_n^{(l,m)} = F^{(l,m)}$ for the homogeneous flight.  We now add an additional segment to each term in this Neumann series using a free-path distribution that is proportional to the attenuation law for leaving a collision.  This creates a ficticious collision density that essentially applies Eq.(\ref{eq:I}) to the in-scattering rate density, which follows our previous approach for the case of isotropic scattering~\cite{deon2019reciprocalii}.  From Eq.(\ref{eq:hseries}) it is clear how a given $h_n$ relates to the previous order $h_{n-1}$, and we find
    \begin{equation}
      h_\phi^{(l)}(u) = \sum_{m=0}^\infty A_m h^{(m)} F_X^{(l,m)}
    \end{equation}
    where the $F$ functions use the attenuation law for leaving a collision (Eq.(\ref{eq:Xc})) instead of a free path distribution,
    \begin{equation}\label{eq:FX}
      F_X^{(l,m)}(u) =  c \int_0^\infty X_c(y) dy \left[ i^m P_m\left( \frac{d}{i d z} \right)\left( j_l(z) \right) \right]_{z \equiv y u}.
    \end{equation}
    This accounts for all flux that arises from collisions in the system. Adding the uncollided flux from the source we find the total fluence
    \begin{equation}\label{eq:fluencesolved}
      \phi(r) = \frac{X_0(s)}{4 \pi r^2} + \frac{1}{4 \pi c} \int_0^\infty h_\phi^{(0)}(u) u j_k(r u) du
    \end{equation}
    where
    \begin{equation}
      X_0(s) = \int_s^\infty p_1(s') ds'.
    \end{equation}
    The full radiance integrated around a given cone with cosine $\mu$ is
    \begin{equation}
      I(r,\mu) = \frac{X_0(s)}{4 \pi r^2}\delta(1-\mu) + \sum_{l=0}^\infty I_l(r) (2l+1) P_l(\mu)
    \end{equation}
    with
    \begin{equation}
      I_l(r) = \frac{1}{8 \pi c} \int_0^\infty h_\phi^{(l)}(u) \, u \, j_l(r \, u) du.
    \end{equation}
    Radiance at some position $\pos$ and direction $\dir$ in the system is given by $I(\pos,\dir) = I(r,\mu) / (2\pi)$.

    \subsubsection{The case of classical exponential random flights}
      For clarity, we briefly examine classical radiative transfer under the present formalism.
      
      In classical linear transport with no spatial correlation between collisions in homogeneous media, the above derivation is equivalent to alternatives involving Kuscer/Chandrasekhar polynomials that satisfy a two-term recurrence~\cite{davison00,kuvsvcer1955milne,grosjean63b,ganapol03}.  With $p_c(s) = \e^{-s}$, the $F$ functions reduce to
      \begin{equation}
        F^{(l,m)}(x) = c \frac{i^{m-l}}{2} \int_{-1}^1 \frac{P_l(\mu) P_m(\mu)}{1-i x \mu} d\mu
      \end{equation}
      with known general solutions in terms of Legendre Q functions~\cite{grosjean63b,vanmassenhove67}, the first few low order terms being
      \begin{equation}
        F^{(0,0)}(u) = c \frac{\tan ^{-1}(u)}{u}, \quad F^{(0,1)}(u) = c \frac{\tan ^{-1}(u)-u}{u^2}, \quad 
        F^{(1,1)}(u) = \frac{c \left(u-\tan ^{-1}(u)\right)}{u^3}.
      \end{equation}

  \section{Gamma random flights in 3D}

    To test Grosjean's derivations for the case of non-exponential random flights we chose gamma random flights~\cite{beghin2010moving,caer11,pogorui2011isotropic,deon14}, which admit explicit solutions in some cases and have a number of interesting properties with respect to diffusion theory.  Intercollision free-path lengths are distributed according the normalized gamma distribution
    \begin{equation}\label{eq:pcgamma}
      p_c(s) = \frac{\e^{s} s^{a-1}}{\Gamma(a)}, \quad a > 0,
    \end{equation}
    which includes classical exponential transport when $a = 1$.  For Monte Carlo validation, random free-path lengths are easily sampled from
    \begin{equation}
      s = -\log \xi_1 \xi_2 ... \xi_a,
    \end{equation}
    where $\xi_n \in [0,1]$ are $a$ independent random uniform variates.  We used the generalized collision estimator for collision-rate density and radiance~\cite{deon2019reciprocalii} to compute the Monte Carlo reference solutions below.

    Combining Eq.(\ref{eq:pcgamma}) with (\ref{eq:F}) we require the integrals
    \begin{equation}
      F^{(l,m)}(u) = \frac{c}{2} \int_0^\infty \int_{-1}^1 \frac{\e^{z} z^{a-1}}{\Gamma(a)} i^{m-l} P_l(\mu ) P_m(\mu ) e^{i \mu  u z} d\mu dz = \frac{c}{2} i^{m-l} \int_{-1}^1  \frac{P_l(\mu ) P_m(\mu )}{(1-i \mu  u)^{a}} d\mu
    \end{equation}
    in the general case $a > 0$.  For the case $m = 0$, we found a general solution
    \begin{equation}
      F^{(l,0)}(u) = \frac{\sqrt{\pi } c 2^{-l-1} u^l \Gamma (a+l)}{\Gamma (a)} \, _2\tilde{F}_1\left(\frac{a+l}{2},\frac{1}{2} (a+l+1);l+\frac{3}{2};-u^2\right)
    \end{equation}
    using the regularized hypergeometric function $_2\tilde{F}_1$.  We suspect a completely general solution is possible using a two-term recurrence, similar to the exponential case, but we did not find it.

    \paragraph{Isotropic Scattering}
    For the case of isotropic scattering, the expansion coefficients in Eq.(\ref{eq:phaseexpansion}) are
    \begin{equation}\label{eq:iso}
      A_0 = 1, \quad A_{l > 0} = 0.
    \end{equation}
    The general solution of the linear system for correlated emission (\ref{eq:hsystem}) with expansion coefficients (\ref{eq:iso}) yields
      \begin{equation}
        h^{(l)} = \frac{2 u F^{(l,0)}}{\pi -\pi  F^{(0,0)}}
      \end{equation}
      and, for the uncorrelated point source, solution of Eq.(\ref{eq:hsystemU}) is
      \begin{equation}
        h^{(l)} = \frac{2 u}{\pi} \left( \frac{F^{(l,0)} F_1^{(0,0)}}{1-F^{(0,0)}} + F_1^{(l,0)} \right).
      \end{equation}

    \paragraph{Linearly-Anisotropic Scattering}

    With linearly-anistropic scattering with parameter $-1 < b < 1$, we find
    \begin{equation}\label{eq:linaniso}
      A_0 = 1, \quad A_1 = b, \quad A_{l > 1} = 0,
    \end{equation}
    yielding expansion functions for correlated emission
      \begin{equation}\label{eq:h0linaniso}
         h^{(l)} = -\frac{2 u ((b F^{(1,1)}(u)-1) F^{(l,0)}(u)+b F^{(0,1)}(u) F^{(l,1)}(u))}{\pi  \left(b \left((F^{(0,1)}(u))^2-F^{(1,1)}(u)\right)+F^{(0,0)}(u) (b
         F^{(1,1)}(u)-1)+1\right)}.
      \end{equation}
      The bulky expressions for the uncorrelated case are omitted.

      \paragraph{Rayleigh Scattering}
        We also consider a simple three-term phase function due to Rayleigh that has application in light scattering~\cite{chandrasekhar60}, with
        \begin{equation}\label{eq:ray}
          A_0 = 1, \quad A_1 = 0, \quad A_2 = \frac{1}{2}, \quad A_{l > 2} = 0
        \end{equation}
        yielding correlated expansion functions
      \begin{equation}\label{eq:h0ray}
        h^{(l)} = \frac{2 u ((2-F^{(2,2)}(u)) F^{(l,0)}(u)+F^{(0,2)}(u) F^{(l,2)}(u))}{\pi  (-(F^{(0,2)}(u))^2+F^{(0,0)}(u) (F^{(2,2)}(u)-2)-F^{(2,2)}(u)+2)}.
      \end{equation}
      The bulky expressions for the uncorrelated case are omitted.

  \subsection{Gamma-2 random flight in 3D}
    With $a = 2$ and an intercollision FPD $p_c(s) = e^{-s} s$, we find, using Eq.(\ref{eq:F}),
    \begin{align}
      &F^{(0,0)}(u) = \frac{c}{1+u^2} , \quad F^{(0,1)}(u) = c \left( \frac{1}{u^3+u}-\frac{\tan ^{-1}(u)}{u^2} \right) , \quad F^{(1,1)}(u) = c \frac{2 \tan ^{-1}(u)-\frac{u \left(u^2+2\right)}{u^2+1}}{u^3} \nonumber \\
      &F^{(0,2)}(u) = c \frac{u \left(\frac{1}{u^2+1}+2\right)-3 \tan ^{-1}(u)}{u^3} , \quad F^{(1,2)}(u) = c \frac{\left(u^2+1\right) \left(u^2+9\right) \tan ^{-1}(u)-u \left(7 u^2+9\right)}{2
      \left(u^6+u^4\right)} \label{eq:F:gamma2} \\
      &F^{(2,2)}(u) = c \frac{u \left(u^2+\frac{1}{u^2+1}+8\right)-3 \left(u^2+3\right) \tan ^{-1}(u)}{u^5} .\nonumber
    \end{align}

    \subsubsection{Linearly-anisotropic scattering}
\begin{figure}
        \centering
        \hspace*{0cm}
        \includegraphics[width=\linewidth]{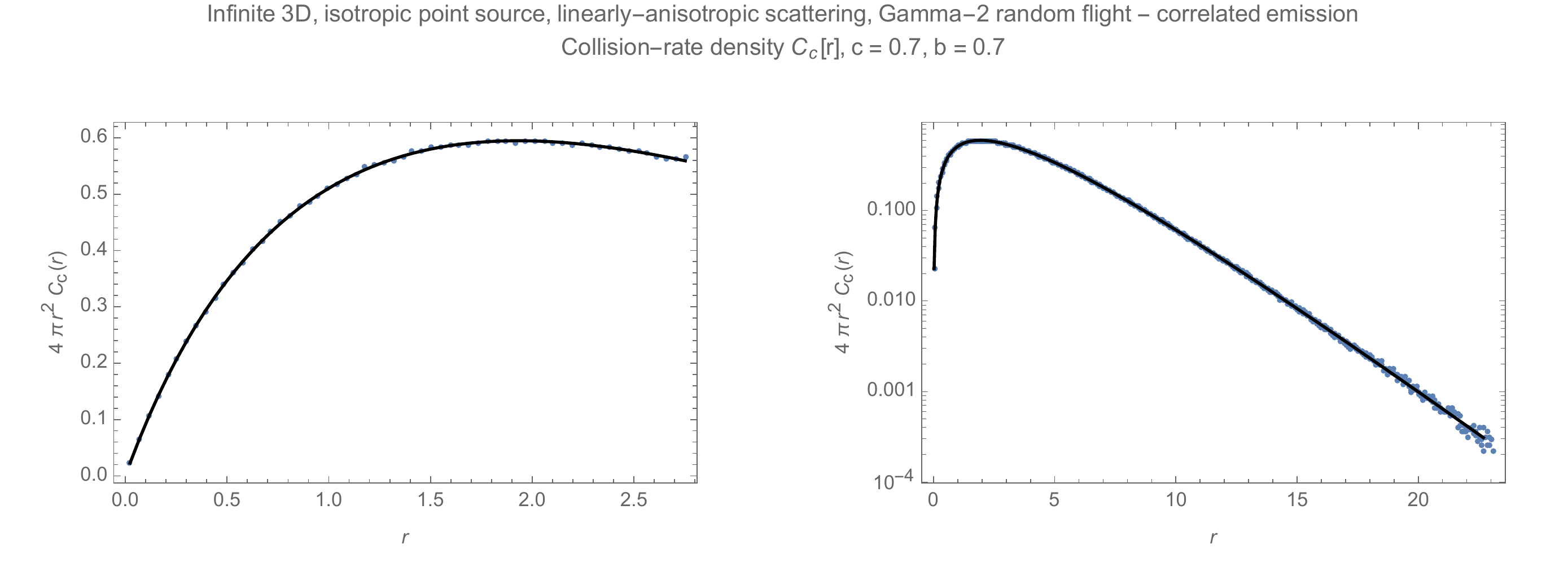}
        \caption{\label{fig:Cc-gamma2-linaniso}Scalar collision-rate density $C_c(r)$ about an isotropic point source in 3D with linearly-anisotropic scattering and intercollision free-path lengths drawn from $e^{-s}s$.  Validation of Eq.(\ref{eq:Cc:gamma2:linaniso}) (continuous) with respect to Monte Carlo (dots).} 
      \end{figure}

      Combining Eq.(\ref{eq:F:gamma2}) with Eq.(\ref{eq:h0linaniso}) we find
      \begin{equation}
        h^{(0)} = \frac{2 c u \left(b c u^2-b c \left(u^2+1\right) \tan ^{-1}(u)^2+u^4\right)}{\pi  \left(b
        c^2 \left(u^2+1\right) \tan ^{-1}(u)^2+b c u^2 \left(-c+u^2+2\right)-2 b c
        \left(u^2+1\right) u \tan ^{-1}(u)+u^4 \left(-c+u^2+1\right)\right)}
      \end{equation}
      yielding scalar collision-rate density
      \begin{equation}\label{eq:Cc:gamma2:linaniso}
        C_c(r) = \frac{1}{2 \pi
        ^2 r} \int_0^\infty \frac{u \left(b c u^2-b c \left(u^2+1\right) \tan ^{-1}(u)^2+u^4\right)}{ \left(u^4 ((b-1) c+1)-b (c-2) c u^2+b c \left(u^2+1\right) \tan ^{-1}(u) \left(c
        \tan ^{-1}(u)-2 u\right)+u^6\right)} \sin (r u) \, du.
      \end{equation} 
      A comparison of this result to Monte Carlo reference is provided in Figure~\ref{fig:Cc-gamma2-linaniso}.  While diffusion is an exact result for collision-rate density in 3D with gamma-2 flights and isotropic scattering~\cite{deon14}, we see that this does not extend to more general phase functions.

\begin{figure}
        \centering
        \hspace*{0cm}
        \includegraphics[width=\linewidth]{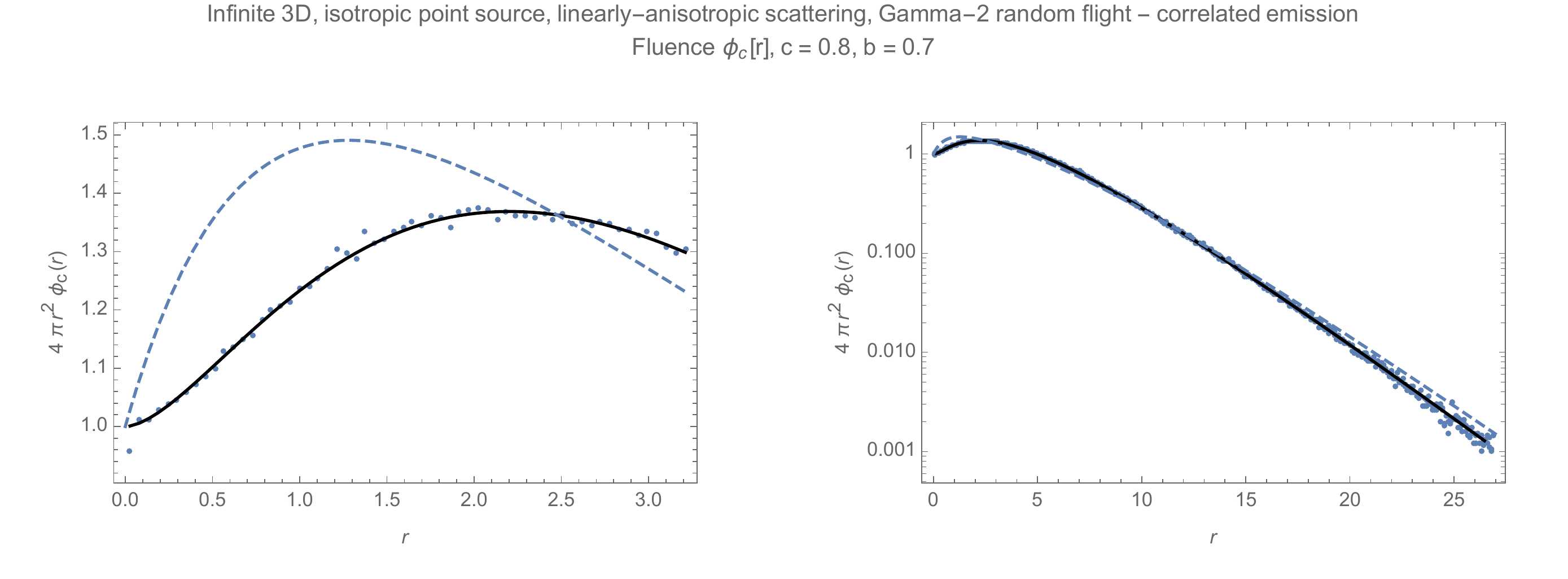}
        \caption{\label{fig:phic-gamma2-linaniso}Scalar flux / fluence $\phi_c(r)$ about a correlated isotropic point source in 3D with linearly-anisotropic scattering and intercollision free-path lengths drawn from $e^{-s}s$.  Validation of Eq.(\ref{eq:gamma2:linaniso:fluence}) (continuous) with respect to Monte Carlo (dots) and comparison of the diffusion approximation, Eq.(\ref{eq:phidiffusion}).} 
      \end{figure}

      For the fluence, we find
      \begin{equation}
        F_X^{(0,0)} = \frac{1}{u^2+1}+\frac{\tan ^{-1}(u)}{u}, \quad F_X^{(0,1)} = -\frac{u}{u^2+1}
      \end{equation}
      and
      \begin{equation}\label{eq:gamma2:linaniso:fluence}
        h_\phi^{(0)} = -\frac{2 c^2 \left(-\left(u^2+1\right) u^2 \tan ^{-1}(u) \left(b
        \left(c-u^2\right)+u^2\right)+b c \left(u^2+1\right) u \tan ^{-1}(u)^2+b c
        \left(u^2+1\right)^2 \tan ^{-1}(u)^3+u^3 \left(-\left(b
        \left(c+u^2\right)+u^2\right)\right)\right)}{\pi  \left(u^2+1\right) \left(b c^2
        \left(u^2+1\right) \tan ^{-1}(u)^2+b c u^2 \left(-c+u^2+2\right)-2 b c
        \left(u^2+1\right) u \tan ^{-1}(u)+u^4 \left(-c+u^2+1\right)\right)}
      \end{equation}
      The fluence then follows from Eq.(\ref{eq:fluencesolved}).  Taking a $(0,2)$ order Pade approximant of the Fourier-transformed density $\pi h_\phi^{(0)}/(2 c u)$ we find the diffusion appromxation for the fluence~\cite{deon14}
      \begin{equation}\label{eq:phidiffusion}
        \phi(r) \approx \frac{e^{-r} (r+1)}{4 \pi  r^2}-\frac{3 c (b c-3) \exp \left(-\frac{\sqrt{3} r}{\sqrt{\frac{b \left(2 c^2-4 c+3\right)-6
        c+15}{(c-1) (b c-3)}}}\right)}{2 \pi  r \left(b \left(2 c^2-4 c+3\right)-6 c+15\right)}
      \end{equation}
      A comparison of these result to Monte Carlo reference is provided in Figure~\ref{fig:phic-gamma2-linaniso}.

\begin{figure}
        \centering
        \hspace*{0cm}
        \includegraphics[width=.4\linewidth]{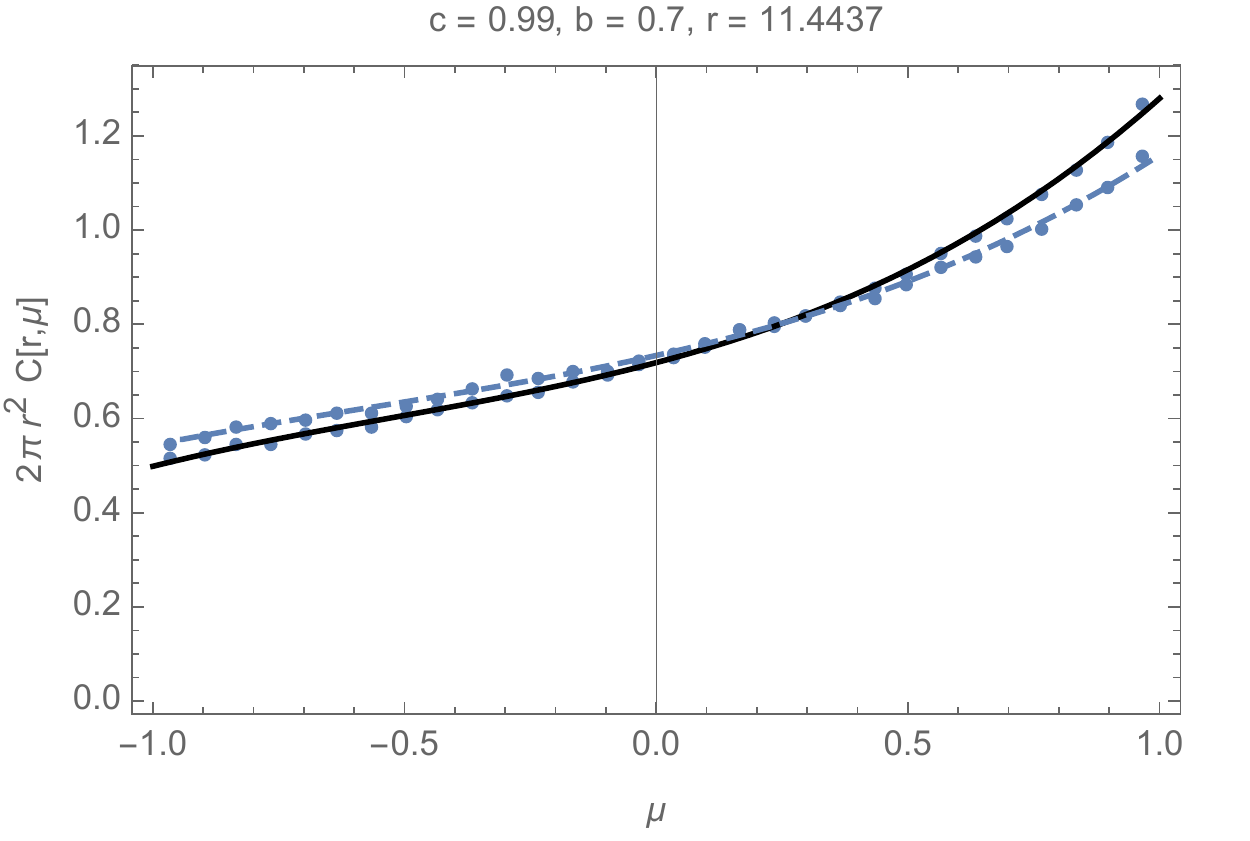}
        \caption{\label{fig:gamma2:angularcompare}Comparison of the angular collision rate density $C(r,\mu)$ (continuous) and the classically scaled radiance $I(r,\mu)/\s$ (dashed) for gamma-2 flights and linearly-anisotropic scattering showing agreement with Monte Carlo (dots) and how the two densities are not proportional in GRT.} 
      \end{figure}

      In Figure~\ref{fig:gamma2:angularcompare} we compare the angular distributions $C(r,\mu)$ and $I(r,\mu) / \s$ using 4 term Legendre expansions about a correlated point source at a radius $r = 11.4437$.  At this distance from the point source this low order expansion seems reasonably accurate with respect to Monte Carlo and illustrates how collision rate and flux are not proportional in GRT.

\begin{figure}
        \centering 
        \hspace*{0cm}
        \includegraphics[width=\linewidth]{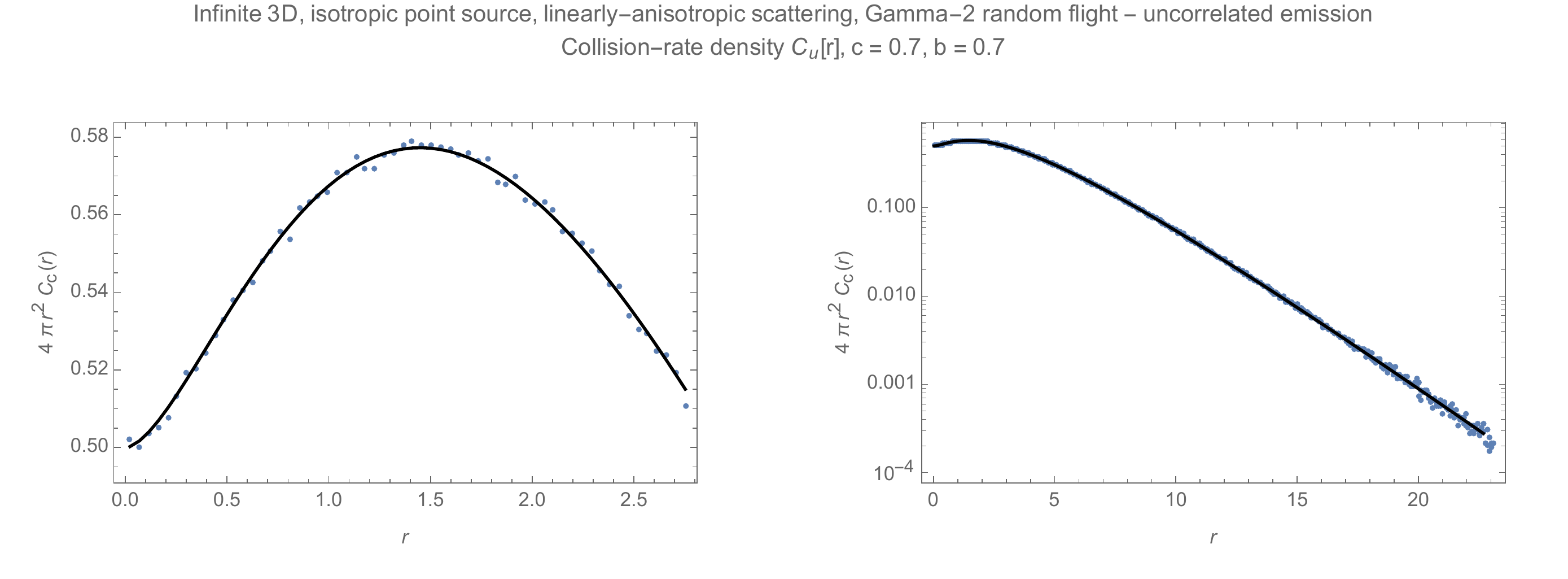}
        \caption{\label{fig:Cu-gamma2-linaniso}Scalar collision-rate density $C_u(r)$ about an uncorrelated-emission isotropic point source in 3D with linearly-anisotropic scattering and intercollision free-path lengths drawn from $e^{-s}s$.  Validation of Eq.(\ref{eq:Cu:gamma2:linaniso}) (continuous) with respect to Monte Carlo (dots).} 
      \end{figure}

      For the uncorrelated source, we find (\ref{eq:F1})
      \begin{equation}
        F_1^{(0,0)}(u) = \frac{c}{2 u^2+2}+\frac{c \tan ^{-1}(u)}{2 u}, \quad F_1^{(0,1)}(u) = -\frac{c u}{2 \left(u^2+1\right)}
      \end{equation}
      giving (\ref{eq:hsystemU})
      \begin{equation}
        h^{(0)} = \frac{c u \left(2 b c u^2-2 b c \left(u^2+1\right) \tan
   ^{-1}(u)^2+u^4+\left(u^5+u^3\right) \tan ^{-1}(u)\right)}{\pi  \left(b c^2
   \left(u^2+1\right) \tan ^{-1}(u)^2+b c u^2 \left(-c+u^2+2\right)-2 b c
   \left(u^2+1\right) u \tan ^{-1}(u)+u^4 \left(-c+u^2+1\right)\right)}
      \end{equation}
      for the scalar collision rate density
      \begin{multline}\label{eq:Cu:gamma2:linaniso}
        C_u(r) = \frac{e^{-r} (r+1)}{8 \pi  r^2} \\+ \int_0^\infty \frac{c \sin (r u) \left(-\left(u^2+1\right) \tan ^{-1}(u) \left(-b c u^2+b c \tan
        ^{-1}(u) \left(\left(u^2+1\right) \tan ^{-1}(u)+u\right)+(b-1) u^4\right)+b u^3
        \left(c+u^2\right)+u^5\right)}{4 \pi ^2 r \left(u^2+1\right) \left(u^4 ((b-1) c+1)-b
        (c-2) c u^2+b c \left(u^2+1\right) \tan ^{-1}(u) \left(c \tan ^{-1}(u)-2
        u\right)+u^6\right)} du.
      \end{multline} 
      A comparison of this result to Monte Carlo reference is provided in Figure~\ref{fig:Cu-gamma2-linaniso}.

    \subsection{Gamma-3 random flight in 3D}
      With intercollision FPD $p_c(s) = \frac{1}{2} e^{-s} s^2$ ($a = 3$) we find
      \begin{align}
        &F^{(0,0)}(u) = \frac{c}{\left(u^2+1\right)^2}, \quad F^{(0,1)}(u) = -\frac{c u}{\left(u^2+1\right)^2} , \quad F^{(1,1)}(u) = \frac{c \left(\frac{2 u^3+u}{\left(u^2+1\right)^2}-\tan ^{-1}(u)\right)}{u^3} \nonumber \\
        &F^{(0,2)}(u) = \frac{1}{2} c \left(\frac{3 \tan ^{-1}(u)}{u^3}+\frac{-5
        u^2-3}{\left(u^3+u\right)^2}\right) , \quad F^{(1,2)}(u) = \frac{3 c \left(u \left(\frac{1}{u^2+1}+2\right)-3 \tan ^{-1}(u)\right)}{2 u^4}-\frac{c
        u}{\left(u^2+1\right)^2} \label{eq:F:gamma3} \\
        &F^{(2,2)}(u) = \frac{c \left(3 \left(u^2+9\right) \tan ^{-1}(u)-\frac{u \left(19 u^4+48
        u^2+27\right)}{\left(u^2+1\right)^2}\right)}{2 u^5}. \nonumber
      \end{align}

      \subsubsection{Linearly-anisotropic scattering}
\begin{figure}
        \centering
        \hspace*{0cm}
        \includegraphics[width=\linewidth]{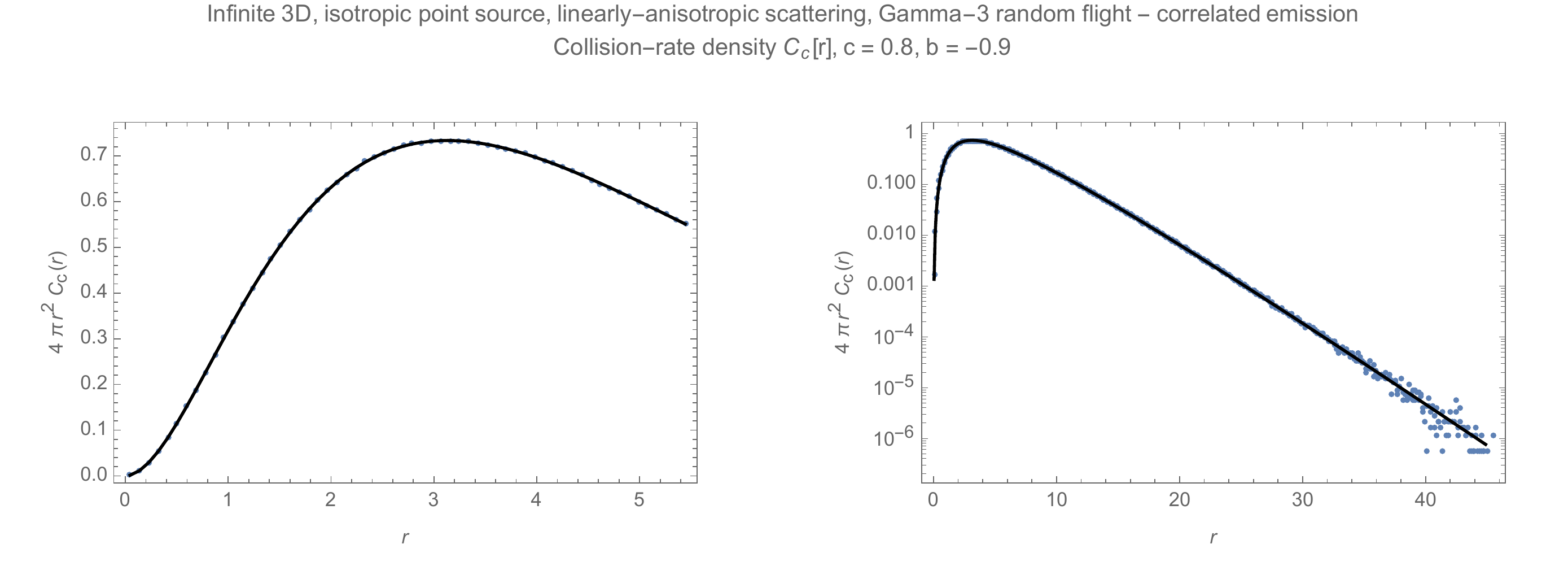}
        \caption{\label{fig:Cu-gamma3-linaniso}Scalar collision-rate density $C_c(r)$ about an isotropic point source in 3D with linearly-anisotropic scattering and intercollision free-path lengths drawn from $e^{-s}s^2/2$.  Validation of Eq.(\ref{eq:C0:gamma3:linaniso}) (continuous) with respect to Monte Carlo (dots).} 
      \end{figure}

        Combining Eq.(\ref{eq:F:gamma3}) with Eq.(\ref{eq:h0linaniso}) we find
      \begin{equation}
        h^{(0)} = \frac{2 c u \left(-b c u+b c \tan ^{-1}(u)+u^3\right)}{\pi  \left(b c u \left(c-2
        u^2-1\right)+b c \left(\left(u^2+1\right)^2-c\right) \tan ^{-1}(u)+u^3
        \left(\left(u^2+1\right)^2-c\right)\right)}
      \end{equation}
      yielding scalar collision-rate density
      \begin{equation}\label{eq:C0:gamma3:linaniso}
        C_c(r) = \frac{1}{2 \pi
        ^2 r} \int_0^\infty \frac{u  \left(-b c u+b c \tan ^{-1}(u)+u^3\right)}{b c u \left(c-2
        u^2-1\right)+b c \left(\left(u^2+1\right)^2-c\right) \tan ^{-1}(u)+u^3
        \left(\left(u^2+1\right)^2-c\right)} \sin (r u) \, du.
      \end{equation} 
      A comparison of this result to Monte Carlo reference is provided in Figure~\ref{fig:Cu-gamma3-linaniso}.

    \subsection{Gamma-4 random flight in 3D}
\begin{figure}
        \centering
        \hspace*{0cm}
        \includegraphics[width=\linewidth]{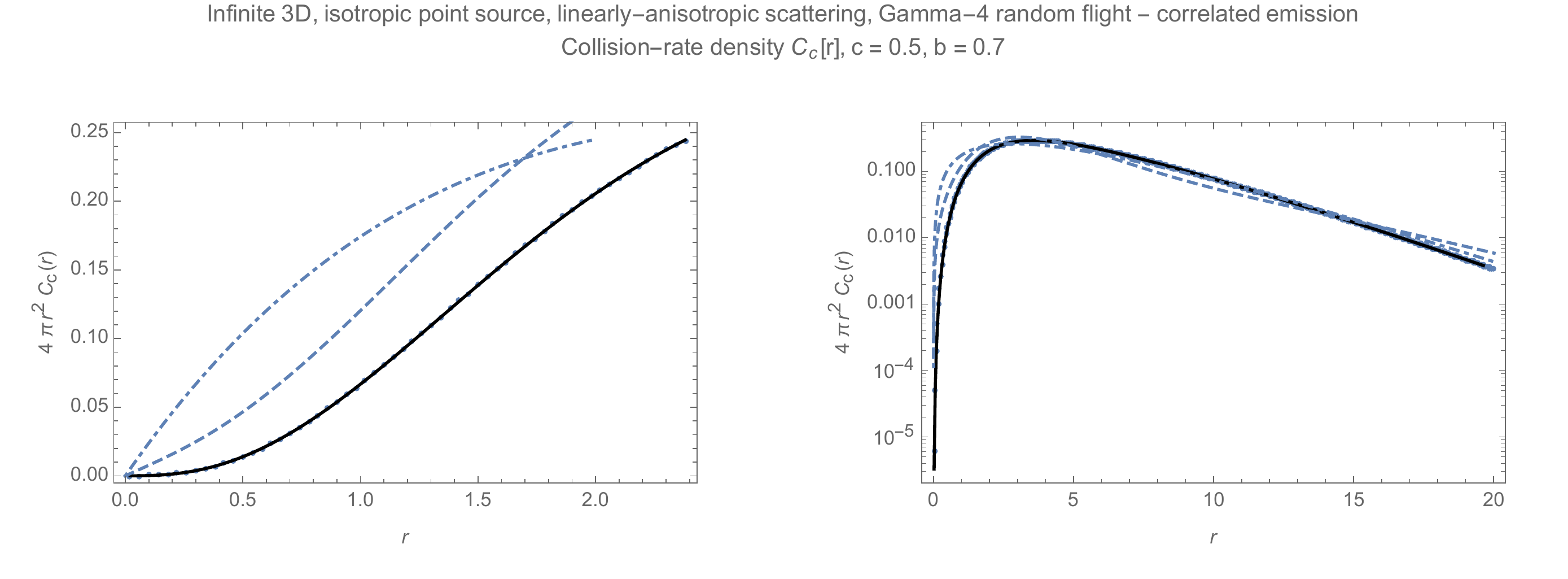}
        \caption{\label{fig:Cu-gamma4-linaniso}Scalar collision-rate density $C_c(r)$ about an isotropic point source in 3D with linearly-anisotropic scattering and intercollision free-path lengths drawn from $\frac{1}{6} e^{-s} s^3$.  Validation of Eq.(\ref{eq:C0:gamma4:linaniso}) (continuous) with respect to Monte Carlo (dots).  Comparisons to a classical diffusion approximation (Eq.(\ref{eq:gamma4linanisoCcClassicalDiffusion}), dot-dashed) and modified-diffusion approximation (Eq.(\ref{eq:gamma4linanisoCcGrosjeanDiffusion}), dashed) are also shown. } 
      \end{figure}

      With $a = 4$ we find that the exact fluence about the point source with linearly-anisotropic scattering can be expressed explicitly as a sum of diffusion modes.  With the intercollision FPD $p_c(s) = \frac{1}{6} e^{-s} s^3$ we find
      \begin{align}
        &F^{(0,0)}(u) = -\frac{c \left(u^2-3\right)}{3 \left(u^2+1\right)^3}, \quad F^{(0,1)}(u) = -\frac{4 c u}{3 \left(u^2+1\right)^3}, \quad F^{(1,1)}(u) = \frac{c-3 c u^2}{3 \left(u^2+1\right)^3}, \quad F^{(0,2)}(u) = \frac{4 c u^2}{3 \left(u^2+1\right)^3} \nonumber \\
        &F^{(1,2)}(u) = \frac{c \left(9 \tan ^{-1}(u)-\frac{u \left(23 u^4+24
        u^2+9\right)}{\left(u^2+1\right)^3}\right)}{6 u^4} , \quad F^{(2,2)}(u) = \frac{c \left(\frac{u \left(11 u^6+60 u^4+72 u^2+27\right)}{\left(u^2+1\right)^3}-27 \tan
        ^{-1}(u)\right)}{3 u^5}. \label{eq:F:gamma4}
      \end{align}
Combining Eq.(\ref{eq:F:gamma4}) with Eq.(\ref{eq:h0linaniso}) we find
      \begin{equation}
        h^{(0)} = -\frac{2 c u \left(b c+u^4-2 u^2-3\right)}{\pi  \left(b c^2+c \left(u^2+1\right) \left(b
        \left(3 u^2-1\right)+u^2-3\right)+3 \left(u^2+1\right)^4\right)}
      \end{equation}
      yielding scalar collision-rate density
      \begin{equation}\label{eq:C0:gamma4:linaniso}
        C_c(r) = \frac{1}{2 \pi
        ^2 r} \int_0^\infty -\frac{u \left(b c+u^4-2 u^2-3\right) }{ \left(b c^2+c
        \left(u^2+1\right) \left(b \left(3 u^2-1\right)+u^2-3\right)+3
        \left(u^2+1\right)^4\right)} \sin (r u) \, du.
      \end{equation} 
      A comparison of this result to Monte Carlo reference is provided in Figure~\ref{fig:Cu-gamma4-linaniso}.  The complete scalar collision rate density can be solved by standard contour integration (\cite{grosjean63b}, pp.73--75), yielding
      \begin{equation}
        C_c(r) = \sum_{v \in v^+}  \frac{\e^{-r v}}{4 \pi r} \frac{\left(1-v^2\right) \left(b c+v^4+2 v^2-3\right)}{2 c \left(b \left(2 c+3
        v^4-3\right)+v^4+4 v^2-5\right)}
      \end{equation}
      where $v^+$ is the set of roots with positive real part of the dispersion equation
      \begin{equation}
        c^2 \left(v^4+10 v^2+1\right)-c \left(11 v^4+26 v^2+11\right) \left(1-v^2\right)^3+10
        \left(1-v^2\right)^8 = 0
      \end{equation}
      for which we found two real and two complex roots in $v^+$.  Figure~\ref{fig:Cu-gamma4-linaniso} also includes comparisons of this exact result to two forms of moment-preserving diffusion approximation found using the methods in~\cite{deon2019reciprocalii}.  For the Classical diffusion approximation, we find
      \begin{equation}\label{eq:gamma4linanisoCcClassicalDiffusion}
        C_c(r) \approx \frac{(3 - b c) e^{-\frac{r}{\sqrt{2} \sqrt{\frac{b c+5}{(c-1) (b c-3)}}}}}{8 \pi  r (b
        c+5)}.
      \end{equation}
      and removing the first-collided portion, we find the Grosjean-form diffusion approximation
      \begin{equation}\label{eq:gamma4linanisoCcGrosjeanDiffusion}
        C_c(r) \approx \frac{e^{-r} r}{24 \pi } + \frac{c}{1-c} \frac{e^{-\frac{r}{v}}}{4 \pi  r v^2}, \quad v = \frac{2 b (5 (c-2) c+8)-30 (c-2)}{3 (c-1) (b c-3)}.
      \end{equation}

    \subsection{Gamma-6 flights in 3D}

    With intercollision FPD $p_c(s) = \frac{1}{120} e^{-s} s^5$ we find
    \begin{align}
      &F^{(0,0)}(u) = \frac{c \left(u^4-10 u^2+5\right)}{5 \left(u^2+1\right)^5}, \quad F^{(0,1)}(u) = \frac{2 c u \left(3 u^2-5\right)}{5 \left(u^2+1\right)^5}, \quad F^{(1,1)}(u) = \frac{c \left(5 u^4-38 u^2+5\right)}{15 \left(u^2+1\right)^5} \nonumber \\
      &F^{(0,2)}(u) = -\frac{2 c u^2 \left(u^2-7\right)}{5 \left(u^2+1\right)^5}, \quad F^{(1,2)}(u) = \frac{4 c u \left(3 u^2-1\right)}{5 \left(u^2+1\right)^5} , \quad F^{(2,2)}(u) = \frac{c \left(1-3 u^2\right)^2}{5 \left(u^2+1\right)^5}. \label{eq:F:gamma6}
    \end{align}

      \subsubsection{Rayleigh scattering}

\begin{figure}
          \centering
          \hspace*{0cm}
          \includegraphics[width=\linewidth]{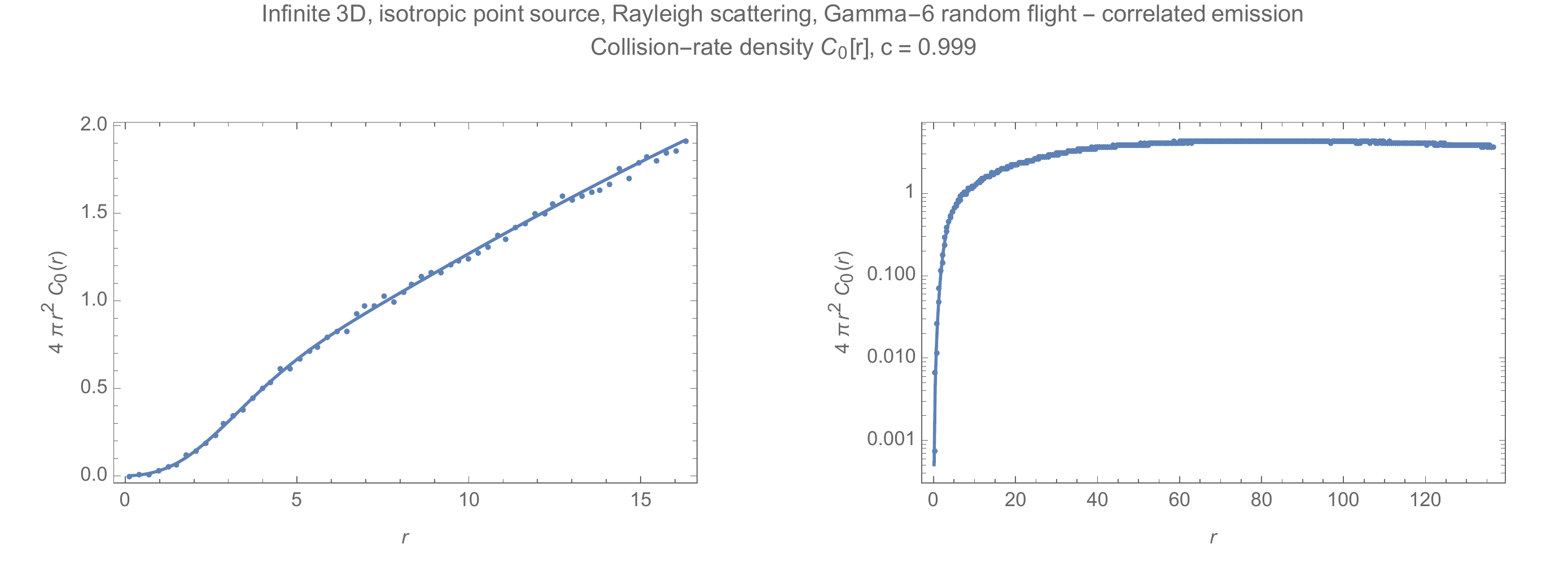}
          \caption{\label{fig:Cu-gamma6-ray}Scalar collision-rate density $C_0(r)$ about an isotropic point source in 3D with Rayleigh scattering and intercollision free-path lengths drawn from $\frac{1}{120} e^{-s} s^5$.  Validation of Eq.(\ref{eq:C0:gamma6:ray}) (continuous) with respect to Monte Carlo (dots).} 
        \end{figure}

        Similar to the gamma-4 case with linearly-anisotropic scattering, with Rayleigh scattering and gamma-6 flights we find a solvable scalar collision rate density.  Combining Eq.(\ref{eq:F:gamma6}) with Eq.(\ref{eq:h0ray}) we find
        \begin{equation}
          h^{(0)} = -\frac{2 c u \left(c \left(u^4-10 u^2+1\right)-2 \left(u^2+1\right)^3 \left(u^4-10
          u^2+5\right)\right)}{\pi  \left(c^2 \left(u^4-10 u^2+1\right)-c \left(11 u^4-26
          u^2+11\right) \left(u^2+1\right)^3+10 \left(u^2+1\right)^8\right)}
        \end{equation}
        yielding scalar collision-rate density
        \begin{equation}\label{eq:C0:gamma6:ray}
          C_c(r) = \frac{1}{2 \pi
          ^2 r} \int_0^\infty -\frac{u^2 \left(c \left(u^4-10 u^2+1\right)-2 \left(u^2+1\right)^3 \left(u^4-10
          u^2+5\right)\right) }{u \left(c^2 \left(u^4-10 u^2+1\right)-c
          \left(11 u^4-26 u^2+11\right) \left(u^2+1\right)^3+10 \left(u^2+1\right)^8\right)} \sin (r u) \, du.
        \end{equation} 
        A comparison of this result to Monte Carlo reference is provided in Figure~\ref{fig:Cu-gamma6-ray}. 
        
        The complete scalar collision rate density is then solved by standard contour integration (\cite{grosjean63b}, pp.73--75)
        \begin{equation}
          C_c(r) = \sum_{v \in v^+} \frac{\left(1-v^2\right) \left(c \left(v^4+10 v^2+1\right)+2 \left(v^4+10 v^2+5\right)
          \left(v^2-1\right)^3\right)}{3 c \left(2 c \left(v^4+12
          v^2+3\right)+\left(v^2+3\right) \left(11 v^2+9\right) \left(v^2-1\right)^3\right)} \frac{\e^{-r v}}{4 \pi r}
        \end{equation}
        where $v^+$ is the set of roots with positive real part of the dispersion equation
        \begin{equation}
          c^2 \left(v^4+10 v^2+1\right)-c \left(11 v^4+26 v^2+11\right) \left(1-v^2\right)^3+10
          \left(1-v^2\right)^8 = 0.
        \end{equation}
        In this case we always noted two real roots and six complex roots for a variety of absorption levels $c$.

      \section{Conclusion}

        We have derived the point source Green's functions for infinite media with anisotropic scattering in non-classical linear transport where the free-path distributions between collisions and attenuation laws are non-exponential.  The general solutions are expressed as Fourier inversions, which were validated numerically using gamma random flights in 3D.  Distinct solutions for both collision rate, and fluence and their angular counterparts were derived and tested using Monte Carlo.  For low integer-order gamma flights and low Legendre orders we found the solutions to be numerically straightforward to manage.  Higher order angular expansions and more challenging free-path distribution such as power-law flights~\cite{davis06} will require more numerical care when dealing with the oscillatory Fourier inversions.  Nevertheless, the provided gamma flight solutions provide important benchmarks for GRT and for validating more efficient approximations, such as SPN~\cite{palmer2020asymptotic}.

      \section{Acknowledgements}

        We thank M.M.R. Williams for helpful feedback on the manuscript and Forrest Brown for bringing several older works~\cite{randall1964,doub1961particle} to our attention.

\bibliographystyle{acmsiggraph}
\bibliography{grtanisogreen} 

\begin{thebibliography}{\protect\citename{Pogorui and Rodr{\'\i}guez-Dagnino
  }2011}

\bibitem[\protect\citename{Alt }1980]{alt1980biased}
{\sc Alt, W.}
\newblock 1980.
\newblock Biased random walk models for chemotaxis and related diffusion
  approximations.
\newblock {\em Journal of mathematical biology 9}, 2, 147--177.
\newblock {\color{gray}{https://doi.org/10.1007/BF00275919}}.

\bibitem[\protect\citename{Audic and Frisch }1993]{audic1993monte}
{\sc Audic, S., and Frisch, H.}
\newblock 1993.
\newblock {Monte-Carlo simulation of a radiative transfer problem in a random
  medium: Application to a binary mixture}.
\newblock {\em Journal of Quantitative Spectroscopy and Radiative Transfer 50},
  2, 127--147.
\newblock {\color{gray}{https://doi.org/10.1016/0022-4073(93)90113-V}}.

\bibitem[\protect\citename{Beghin and Orsingher }2010]{beghin2010moving}
{\sc Beghin, L., and Orsingher, E.}
\newblock 2010.
\newblock Moving randomly amid scattered obstacles.
\newblock {\em Stochastics: An International Journal of Probability and
  Stochastics Processes 82}, 2, 201--229.
\newblock {\color{gray}{https://doi.org/10.1080/17442500903359163}}.

\bibitem[\protect\citename{Binzoni et~al\mbox{.} }2018]{binzoni2018generalized}
{\sc Binzoni, T., Martelli, F., and Kozubowski, T.~J.}
\newblock 2018.
\newblock Generalized time-independent correlation transport equation with
  static background: influence of anomalous transport on the field
  autocorrelation function.
\newblock {\em JOSA A 35}, 6, 895--902.
\newblock {\color{gray}{https://doi.org/10.1364/JOSAA.35.000895}}.

\bibitem[\protect\citename{Bitterli et~al\mbox{.} }2018]{bitterli2018radiative}
{\sc Bitterli, B., Ravichandran, S., M{\"u}ller, T., Wrenninge, M., Nov{\'a}k,
  J., Marschner, S., and Jarosz, W.}
\newblock 2018.
\newblock A radiative transfer framework for non-exponential media.
\newblock {\em ACM Transactions on Graphics 37}, 6.
\newblock {\color{gray}{https://doi.org/10.1145/3272127.3275103}}.

\bibitem[\protect\citename{Burrus }1960]{burrus1960radiation}
{\sc Burrus, W.}
\newblock 1960.
\newblock Radiation transmission through boral and similar heterogeneous
  materials consisting of randomly distributed absorbing chunks.
\newblock Tech. rep., Oak Ridge National Lab., Tenn.
\newblock {\color{gray}{https://doi.org/10.2172/4196641}}.

\bibitem[\protect\citename{Case et~al\mbox{.} }1953]{case53}
{\sc Case, K.~M., de~Hoffman, F., and Placzek, G.}
\newblock 1953.
\newblock {\em Introduction to the Theory of Neutron Diffusion}, vol.~1.
\newblock US Government Printing Office.

\bibitem[\protect\citename{Chandrasekhar }1960]{chandrasekhar60}
{\sc Chandrasekhar, S.}
\newblock 1960.
\newblock {\em Radiative Transfer}.
\newblock Dover.

\bibitem[\protect\citename{Davis and Marshak }2004]{davis04}
{\sc Davis, A.~B., and Marshak, A.}
\newblock 2004.
\newblock Photon propagation in heterogeneous optical media with spatial
  correlations: enhanced mean-free-paths and wider-than-exponential free-path
  distributions.
\newblock {\em Journal of Quantitative Spectroscopy and Radiative Transfer 84},
  1, 3--34.
\newblock {\color{gray}{https://doi.org/10.1016/S0022-4073(03)00114-6}}.

\bibitem[\protect\citename{Davis and Xu }2014]{davis14}
{\sc Davis, A.~B., and Xu, F.}
\newblock 2014.
\newblock A generalized linear transport model for spatially correlated
  stochastic media.
\newblock {\em Journal of Computational and Theoretical Transport 43}, 1-7,
  474--514.
\newblock {\color{gray}{https://doi.org/10.1080/23324309.2014.978083}}.

\bibitem[\protect\citename{Davis }2006]{davis06}
{\sc Davis, A.~B.}
\newblock 2006.
\newblock Effective propagation kernels in structured media with broad spatial
  correlations, illustration with large-scale transport of solar photons
  through cloudy atmospheres.
\newblock In {\em Computational Methods in Transport}. Springer, 85--140.
\newblock {\color{gray}{https://doi.org/10.1007/3-540-28125-8\_5}}.

\bibitem[\protect\citename{Davison }1957]{davison57}
{\sc Davison, B.}
\newblock 1957.
\newblock {\em {Neutron Transport Theory}}.
\newblock Oxford University Press.

\bibitem[\protect\citename{Davison }2000]{davison00}
{\sc Davison, B.}
\newblock 2000.
\newblock {Angular distribution due to an isotropic point source and
  spherically symmetrical eigensolutions of the transport equation (MT-112)}.
\newblock {\em Progress in Nuclear Energy 36}, 3, 323 -- 365.
\newblock Nuclear Reactor Theory in Canada 1943-1946.
\newblock {\color{gray}{https://doi.org/10.1016/S0149-1970(00)00012-3}}.

\bibitem[\protect\citename{d'Eon }2013]{deon14}
{\sc d'Eon, E.}
\newblock 2013.
\newblock {Rigorous Asymptotic and Moment-Preserving Diffusion Approximations
  for Generalized Linear Boltzmann Transport in Arbitrary Dimension}.
\newblock {\em Transport Theory and Statistical Physics 42}, 6-7, 237--297.
\newblock {\color{gray}{https://doi.org/10.1080/00411450.2014.910231}}.

\bibitem[\protect\citename{d'Eon }2018]{deon2018reciprocal}
{\sc d'Eon, E.}
\newblock 2018.
\newblock {A reciprocal formulation of nonexponential radiative transfer. 1:
  Sketch and motivation}.
\newblock {\em Journal of Computational and Theoretical Transport\/}.
\newblock {\color{gray}{https://doi.org/10.1080/23324309.2018.1481433}}.

\bibitem[\protect\citename{d'Eon }2019]{deon2019reciprocalii}
{\sc d'Eon, E.}
\newblock 2019.
\newblock {A reciprocal formulation of nonexponential radiative transfer. 2:
  Monte-Carlo Estimation and Diffusion Approximation}.
\newblock {\em Journal of Computational and Theoretical Transport 48}, 6,
  201--262.
\newblock {\color{gray}{https://doi.org/10.1080/23324309.2019.1677717}}.

\bibitem[\protect\citename{Doub }1961]{doub1961particle}
{\sc Doub, W.}
\newblock 1961.
\newblock Particle self-shielding in plates loaded with spherical poison
  particles.
\newblock {\em Nuclear Science and Engineering 10}, 4, 299--307.
\newblock {\color{gray}{https://doi.org/10.13182/NSE61-A15371}}.

\bibitem[\protect\citename{Feller }1971]{feller1971introduction}
{\sc Feller, W.}
\newblock 1971.
\newblock {\em An Introduction to Probability theory and its application Vol
  II}.
\newblock John Wiley and Sons.

\bibitem[\protect\citename{Frank and Sun }2018]{frank2018fractional}
{\sc Frank, M., and Sun, W.}
\newblock 2018.
\newblock Fractional diffusion limits of non-classical transport equations.
\newblock {\em Kinetic \& Related Models 11}, 6, 1503--1526.
\newblock {\color{gray}{https://doi.org/10.3934/krm.2018059}}.

\bibitem[\protect\citename{Frank et~al\mbox{.} }2010]{frank10}
{\sc Frank, M., Goudon, T., et~al.}
\newblock 2010.
\newblock {On a generalized Boltzmann equation for non-classical particle
  transport}.
\newblock {\em Kinetic and Related Models 3\/}, 395--407.
\newblock {\color{gray}{https://doi.org/10.3934/krm.2010.3.395}}.

\bibitem[\protect\citename{Frank et~al\mbox{.} }2015]{frank15}
{\sc Frank, M., Krycki, K., Larsen, E.~W., and Vasques, R.}
\newblock 2015.
\newblock {The nonclassical Boltzmann equation and diffusion-based
  approximations to the Boltzmann equation}.
\newblock {\em SIAM Journal on Applied Mathematics 75}, 3, 1329--1345.
\newblock {\color{gray}{https://doi.org/10.1137/140999451}}.

\bibitem[\protect\citename{Ganapol }2003]{ganapol03}
{\sc Ganapol, B.~D.}
\newblock 2003.
\newblock Fourier transform transport solutions in spherical geometry.
\newblock {\em Transport Theory and Statistical Physics 32}, 5, 587 -- 605.
\newblock {\color{gray}{https://doi.org/10.1081/TT-120025067}}.

\bibitem[\protect\citename{Ganapol }2008]{ganapol08}
{\sc Ganapol, B.}, 2008.
\newblock {Analytical Benchmarks for Nuclear Engineering Applications}.

\bibitem[\protect\citename{Grosjean }1951]{grosjean51}
{\sc Grosjean, C.}
\newblock 1951.
\newblock {The Exact Mathematical Theory of Multiple Scattering of Particles in
  an Infinite Medium}.
\newblock {\em Memoirs Kon. Vl. Ac. Wetensch. 13}, 36.

\bibitem[\protect\citename{Grosjean }1963]{grosjean63b}
{\sc Grosjean, C.~C.}
\newblock 1963.
\newblock {A new approximate one-velocity theory for treating both isotropic
  and anisotropic multiple scattering problems. Part I. Infinite homogeneous
  scattering media}.
\newblock Tech. rep., Universiteit, Ghent.

\bibitem[\protect\citename{Ivanov }1994]{ivanov1994resolvent}
{\sc Ivanov, V.}
\newblock 1994.
\newblock Resolvent method: exact solutions of half-space transport problems by
  elementary means.
\newblock {\em Astronomy and Astrophysics 286\/}, 328--337.

\bibitem[\protect\citename{Jarabo et~al\mbox{.} }2018]{jarabo18}
{\sc Jarabo, A., Aliaga, C., and Gutierrez, D.}
\newblock 2018.
\newblock A radiative transfer framework for spatially-correlated materials.
\newblock {\em ACM Transactions on Graphics 37}, 4, 14.
\newblock {\color{gray}{https://doi.org/10.1145/3197517.3201282}}.

\bibitem[\protect\citename{Kostinski }2001]{kostinski01}
{\sc Kostinski, A.~B.}
\newblock 2001.
\newblock On the extinction of radiation by a homogeneous but spatially
  correlated random medium.
\newblock {\em JOSA A 18}, 8, 1929--1933.
\newblock {\color{gray}{https://doi.org/10.1364/JOSAA.18.001929}}.

\bibitem[\protect\citename{Ku{\v{s}}{\v{c}}er }1955]{kuvsvcer1955milne}
{\sc Ku{\v{s}}{\v{c}}er, I.}
\newblock 1955.
\newblock Milne's problem for anisotropic scattering.
\newblock {\em Journal of Mathematics and Physics 34}, 1-4, 256--266.
\newblock {\color{gray}{https://doi.org/10.1002/sapm1955341256}}.

\bibitem[\protect\citename{Larsen and Vasques }2011]{larsen11}
{\sc Larsen, E.~W., and Vasques, R.}
\newblock 2011.
\newblock {A generalized linear Boltzmann equation for non-classical particle
  transport}.
\newblock {\em Journal of Quantitative Spectroscopy and Radiative Transfer
  112}, 4, 619--631.
\newblock {\color{gray}{https://doi.org/10.1016/j.jqsrt.2010.07.003}}.

\bibitem[\protect\citename{Le~Ca{\"e}r }2011]{caer11}
{\sc Le~Ca{\"e}r, G.}
\newblock 2011.
\newblock A new family of solvable pearson-dirichlet random walks.
\newblock {\em Journal of Statistical Physics 144}, 1, 23--45.
\newblock {\color{gray}{https://doi.org/10.1007/s10955-011-0245-4}}.

\bibitem[\protect\citename{Liemert and Kienle }2017]{liemert2017radiative}
{\sc Liemert, A., and Kienle, A.}
\newblock 2017.
\newblock {Radiative transport equation for the Mittag-Leffler path length
  distribution}.
\newblock {\em Journal of Mathematical Physics 58}, 5, 053511.
\newblock {\color{gray}{https://doi.org/10.1063/1.4983682}}.

\bibitem[\protect\citename{Moon et~al\mbox{.} }2007]{moon07}
{\sc Moon, J., Walter, B., and Marschner, S.}
\newblock 2007.
\newblock Rendering discrete random media using precomputed scattering
  solutions.
\newblock {\em Rendering Techniques 2007\/}, 231--242.
\newblock {\color{gray}{https://doi.org/10.2312/EGWR/EGSR07/231-242}}.

\bibitem[\protect\citename{Narasimhan and Nayar }2003]{narasimhan2003shedding}
{\sc Narasimhan, S.~G., and Nayar, S.~K.}
\newblock 2003.
\newblock Shedding light on the weather.
\newblock In {\em 2003 IEEE Computer Society Conference on Computer Vision and
  Pattern Recognition, 2003. Proceedings.}, vol.~1, IEEE, I--I.

\bibitem[\protect\citename{Paasschens }1997]{paasschens97}
{\sc Paasschens, J. C.~J.}
\newblock 1997.
\newblock Solution of the time-dependent boltzmann equation.
\newblock {\em Phys. Rev. E 56}, 1 (Jul), 1135--1141.
\newblock {\color{gray}{https://doi.org/10.1103/PhysRevE.56.1135}}.

\bibitem[\protect\citename{Palmer and Vasques }2020]{palmer2020asymptotic}
{\sc Palmer, R.~K., and Vasques, R.}
\newblock 2020.
\newblock {Asymptotic Derivation of the Simplified $P_N$ Equations for
  Nonclassical Transport with Anisotropic Scattering}.
\newblock {\color{gray}{https://arxiv.org/abs/2001.05890}}.

\bibitem[\protect\citename{Pogorui and Rodr{\'\i}guez-Dagnino
  }2011]{pogorui2011isotropic}
{\sc Pogorui, A.~A., and Rodr{\'\i}guez-Dagnino, R.~M.}
\newblock 2011.
\newblock {Isotropic random motion at finite speed with K-Erlang distributed
  direction alternations}.
\newblock {\em Journal of Statistical Physics 145}, 1, 102.
\newblock {\color{gray}{\\https://doi.org/10.1007/s10955-011-0328-2}}.

\bibitem[\protect\citename{Randall }1964]{randall1964}
{\sc Randall, C.}
\newblock 1964.
\newblock Generalized treatment of particle self-shielding.
\newblock In {\em The Naval Reactors Handbook Vol. 1: Selected Basic
  Techniques}, A.~Radkowsky, Ed. United States Atomic Energy Comission, 553.

\bibitem[\protect\citename{Rukolaine }2016]{rukolaine2016generalized}
{\sc Rukolaine, S.~A.}
\newblock 2016.
\newblock Generalized linear boltzmann equation, describing non-classical
  particle transport, and related asymptotic solutions for small mean free
  paths.
\newblock {\em Physica A: Statistical Mechanics and its Applications 450\/},
  205--216.
\newblock {\color{gray}{https://doi.org/10.1016/j.physa.2015.12.105}}.

\bibitem[\protect\citename{Rybicki }1965]{rybicki1965transfer}
{\sc Rybicki, G.~B.}
\newblock 1965.
\newblock Transfer of radiation in stochastic media.
\newblock Tech. Rep. 180, Smithsonian Astrophysical Observatory, June.

\bibitem[\protect\citename{Sahni }1989]{sahni1989equivalence}
{\sc Sahni, D.}
\newblock 1989.
\newblock Equivalence of generic equation method and the phenomenological model
  for linear transport problems in a two-state random scattering medium.
\newblock {\em Journal of mathematical physics 30}, 7, 1554--1559.
\newblock {\color{gray}{https://doi.org/10.1063/1.528288}}.

\bibitem[\protect\citename{Taine et~al\mbox{.} }2010]{taine2010generalized}
{\sc Taine, J., Bellet, F., Leroy, V., and Iacona, E.}
\newblock 2010.
\newblock Generalized radiative transfer equation for porous medium upscaling:
  Application to the radiative fourier law.
\newblock {\em International Journal of Heat and Mass Transfer 53}, 19-20,
  4071--4081.
\newblock
  {\color{gray}{https://doi.org/10.1016/j.ijheatmasstransfer.2010.05.027}}.

\bibitem[\protect\citename{Tessendorf }2011]{tessendorf11}
{\sc Tessendorf, J.}
\newblock 2011.
\newblock Angular smoothing and spatial diffusion from the {F}eynman path
  integral representation of radiative transfer.
\newblock {\em Journal of Quantitative Spectroscopy and Radiative Transfer
  112}, 4, 751--760.
\newblock {\color{gray}{https://doi.org/10.1016/j.jqsrt.2010.11.004}}.

\bibitem[\protect\citename{Torquato and Lu }1993]{torquato93}
{\sc Torquato, S., and Lu, B.}
\newblock 1993.
\newblock Chord-length distribution function for two-phase random media.
\newblock {\em Physical Review E 47}, 4, 2950.
\newblock {\color{gray}{\\https://doi.org/10.1016/0306-4549(92)90013-2}}.

\bibitem[\protect\citename{Tunaley }1974]{tunaley1974theory}
{\sc Tunaley, J.}
\newblock 1974.
\newblock Theory of ac conductivity based on random walks.
\newblock {\em Physical Review Letters 33}, 17, 1037.
\newblock {\color{gray}{https://doi.org/10.1103/PhysRevLett.33.1037}}.

\bibitem[\protect\citename{Tunaley }1976]{tunaley1976moments}
{\sc Tunaley, J.}
\newblock 1976.
\newblock Moments of the montroll-weiss continuous-time random walk for
  arbitrary starting time.
\newblock {\em Journal of Statistical Physics 14}, 5, 461--463.
\newblock {\color{gray}{https://doi.org/10.1007/BF01040704}}.

\bibitem[\protect\citename{Vanmassenhove and Grosjean }1967]{vanmassenhove67}
{\sc Vanmassenhove, F., and Grosjean, C.}
\newblock 1967.
\newblock {Electromagnetic Scattering}.
\newblock In {\em Proc. Second Interdisciplinary Conf. Electromagnetic
  Scattering}, 721--763.

\bibitem[\protect\citename{Vasques and Larsen }2014]{vasques13}
{\sc Vasques, R., and Larsen, E.~W.}
\newblock 2014.
\newblock Non-classical particle transport with angular-dependent path-length
  distributions. i: Theory.
\newblock {\em Annals of Nuclear Energy 70\/}, 292--300.
\newblock {\color{gray}{https://doi.org/10.1016/j.anucene.2013.12.021}}.

\bibitem[\protect\citename{Wallace }1948]{wallace1948angular}
{\sc Wallace, P.}
\newblock 1948.
\newblock Angular distribution of neutrons inside a scattering and absorbing
  medium.
\newblock {\em Canadian journal of research 26}, 2, 99--114.
\newblock {\color{gray}{https://doi.org/10.1139/cjr48a-011}}.

\bibitem[\protect\citename{Weiss and Rubin }1983]{weissrubin1983random}
{\sc Weiss, G.~H., and Rubin, R.~J.}
\newblock 1983.
\newblock Random walks: theory and selected applications.
\newblock {\em Adv. Chem. Phys 52\/}, 363--505.
\newblock {\color{gray}{https://doi.org/10.1002/9780470142769.ch5}}.

\bibitem[\protect\citename{Williams }1977]{williams1977role}
{\sc Williams, M.}
\newblock 1977.
\newblock On the role of the adjoint boltzmann equation in the calculation of
  energy deposition.
\newblock {\em Journal of Physics D: Applied Physics 10}, 17, 2343.
\newblock {\color{gray}{https://doi.org/10.1088/0022-3727/10/17/006}}.

\bibitem[\protect\citename{Wrenninge et~al\mbox{.} }2017]{wrenninge17}
{\sc Wrenninge, M., Villemin, R., and Hery, C.}
\newblock 2017.
\newblock Path traced subsurface scattering using anisotropic phase functions
  and non-exponential free flights.
\newblock Tech. Rep. 17-07, Pixar.
\newblock
  {\color{gray}{https://graphics.pixar.com/library/PathTracedSubsurface}}.

\bibitem[\protect\citename{Xu et~al\mbox{.} }2016]{xu16}
{\sc Xu, F., Davis, A.~B., and Diner, D.~J.}
\newblock 2016.
\newblock Markov chain formalism for generalized radiative transfer in a
  plane-parallel medium, accounting for polarization.
\newblock {\em Journal of Quantitative Spectroscopy and Radiative Transfer
  184\/}, 14--26.
\newblock {\color{gray}{https://doi.org/10.1016/j.jqsrt.2016.06.004}}.

\bibitem[\protect\citename{Zarrouati et~al\mbox{.}
  }2013]{zarrouati2013statistical}
{\sc Zarrouati, M., Enguehard, F., and Taine, J.}
\newblock 2013.
\newblock Statistical characterization of near-wall radiative properties of a
  statistically non-homogeneous and anisotropic porous medium.
\newblock {\em International journal of heat and mass transfer 67\/}, 776--783.
\newblock
  {\color{gray}{https://doi.org/10.1016/j.ijheatmasstransfer.2013.08.021}}.

\bibitem[\protect\citename{Zoia et~al\mbox{.} }2011]{zoia11e}
{\sc Zoia, A., Dumonteil, E., and Mazzolo, A.}
\newblock 2011.
\newblock Collision densities and mean residence times for d-dimensional
  exponential flights.
\newblock {\em Physical Review E 83}, 4, 041137.
\newblock {\color{gray}{https://doi.org/10.1103/PhysRevE.83.041137}}.

\end{thebibliography}

\end{document}